\documentclass{ws-procs975x65}
\usepackage{graphics}
\newcommand{\be}{\begin{equation}}
\newcommand{\ee}{\end{equation}}
\newcommand{\bea}{\begin{eqnarray}}
\newcommand{\eea}{\end{eqnarray}}

\newcommand{\non}{\nonumber}

\newcommand{\mev}{\,{\rm MeV}}

\begin{document}

\title{Baryon Spectroscopy and the Constituent Quark Model}

\author{A.~W.~THOMAS and R.~D.~YOUNG} 

\address{Thomas Jefferson National Accelerator Facility \\
12000 Jefferson Ave., Newport News VA 23185 USA \\} 

\maketitle

\abstracts{We explore further the idea that the lattice QCD data for
hadron properties in the region $m_\pi^2 > 0.2 {\rm GeV}^2$ can be
described by the constituent quark model. This leads to a natural
explanation of the fact that nucleon excited states are generally
stable for pion masses greater than their physical excitation
energies. Finally we apply these same ideas to the problem of how
pentaquarks might behave in lattice QCD, with interesting conclusions.
}

\section{Introduction}

Studies of lattice QCD have revealed a transition in the behavior of
hadron properties in the region of quark mass $m_\pi \sim 0.4-0.5$
GeV. Beyond this point hadron properties exhibit smooth, slowly
varying behavior as a function of quark mass. The constituent quark
mass is expected to depend linearly on the current quark mass, with $M
= M_0 + c m_q$ and $c \sim 1$. In addition, in a constituent quark
model (CQM) hadron masses are roughly linear in the number of
constituent quarks ($ \approx n_H M$, with $n_H$ the number of
constituent quarks), magnetic moments are proportional to $1/M$ and so
on. Since this is consistent with what is observed it is clear that
the lattice simulations are qualitatively consistent with the
constituent quark picture in the region $m_\pi > 0.5$ GeV.

On the other hand, in the region where the pion mass approaches zero,
we know on model-independent grounds that {\it all} hadron properties
exhibit rapidly varying, non-analytic behavior as a function of $m_q$
-- as a consequence of the pion loops resulting from spontaneous chiral
symmetry breaking~\cite{Li:1971vr}. These loops can change physical
properties by up to 50\% from the predictions based on naive chiral
extrapolation~\cite{Detmold:2001jb}. Since precise lattice
computations at sufficiently low quark mass to see this curvature
unambiguously are not yet possible~\footnote{One possible exception,
which relies on the non-unitary nature of quenched QCD, is the
comparison between N and $\Delta$ properties in quenched
QCD~\cite{Young:2002cj,Young:2003ns}.}, the accurate determination of
physical hadron properties is not possible unless we know the
corresponding chiral coefficients. Thus, while nucleon properties such
as mass, magnetic moments, charge radii and moments of parton
distribution functions can be extracted by chiral extrapolation,
hadron spectroscopy presents a serious challenge. In the case of
pentaquarks, since if they exist their structure is completely
unknown, nothing is known about their chiral behavior. Hence, even if
one finds a signal at large $m_q$, the ambiguity in its physical mass
is hundreds of MeV.

In this report we shall concentrate on the mass region in which constituent 
quark behavior dominates. We aim to understand the lattice data which has 
been obtained so far in terms of a constituent quark picture -- with 
remarkable success.

\section{Access Quark Model -- AccessQM}

In spite of the obvious, qualitative success of the CQM in describing
the quark mass dependence of hadron properties calculated in lattice
QCD in the currently accessible mass region, there has so far been
only one quantitative application of the CQM to it! This approach,
labelled AccessQM by the authors~\cite{Cloet:2002eg}, involved the
magnetic moments of the baryon octet.  The procedure was to use the
simplest CQM to give the octet moments in the mass region where $m_u
\sim m_d \sim m_s$ and then to extrapolate to the region of
physical, light quark masses using the correct chiral coefficients.
The result was in excellent agreement with the experimental
data. Further improvements in the CQM should be investigated in order
to see just how good the agreement can be made.

On a more general level, this success suggests a novel approach to the
CQM, where it would be developed to match lattice QCD data at
relatively large quark mass, rather than trying to match it to
experimental data where one knows that pion corrections can be very
large. In addition, this work suggests that a comparison with lattice
data at larger light quark mass would serve as a useful test of {\it
any} model of hadron structure~\cite{Detmold:2001hq}. 
So far only the chiral quark soliton
model and the cloudy bag model have taken up this
challenge~\cite{Leinweber:1998ej,Leinweber:1999ig,Goeke:2005fs}.

\subsection{N and $\Delta$ Masses}
As the next check on the physical consistency of the CQM at large
quark mass we compare the data for N and $\Delta$ masses from full QCD
with expectations in AccessQM. In Ref.~\cite{Cloet:2002eg} the
constituent quark mass was written as
\be
M = 0.421 + 0.301 m_\pi^2
\label{eq:1}
\ee
with $m_\pi$ in GeV. In a careful comparison of the N and $\Delta$
masses in full and quenched QCD, Young {\it et al.}~\cite{Young:2002cj}
found that the masses in full QCD (from the MILC
Collaboration~\cite{Bernard:2001av}) were well fit by:
\bea
m_\Delta &=& 1.43(3) + 0.75 (8) m_\pi^2  + \Sigma_\Delta( {\rm chiral \, loops}) \non \\
m_{\rm N} &=& 1.24(2) + 0.92 (5) m_\pi^2 + \Sigma_N( {\rm chiral \, loops}) \, ,
\label{eq:2}
\eea
where the chiral loops were calculated using a dipole form factor at the 
baryon-pion vertices with mass parameter 0.8 GeV. In the region where the 
pion mass is large these chiral loops are small and vary slowly. Thus it 
makes sense to compare the slope of the average of the N and $\Delta$ masses 
in that region, namely $0.83 \pm 0.09$ GeV$^{-1}$, with three times the 
coefficient of $m_\pi^2$ in Eq.(\ref{eq:1}), that is 0.90 GeV $^{-1}$. In 
terms of the absolute values, when the light quark mass is comparable with the 
strange quark mass, that is when $m_\pi^2 \sim 0.48\,{\rm GeV}^2$, 
$3M$ is roughly
1.68 GeV, while the average of the N and $\Delta$ masses (from the MILC data) 
is of order 1.60 GeV. Clearly the CQM does a very reasonable job of describing 
the data quantitatively in the large mass region. A more sophisticated 
treatment would also model the hyperfine interaction associated with (say) 
gluon exchange with a term proportional to $1/M$.

\subsection{Unstable Hadrons -- $\Delta(1232)$}
Now we compare the behavior of the mass of the $\Delta(1232)$ resonance with 
the mass of the corresponding decay channel, namely a pion and a nucleon.
Provided that the self-energies of the N and $\Delta$ associated with 
chiral loops are either nearly equal or small, we can write the difference 
between the $\Delta$ mass and that of the threshold for the open N $\pi$ 
channel as:
\be
m_{\Delta} - m_N - m_\pi = 0.19 - 0.17 m_\pi^2 -m_\pi \, .
\label{eq:3}
\ee
This would vanish and hence the $\Delta$ would become bound at $m_\pi
\sim 0.18$ GeV. However, at such a low mass the self-energies 
associated with chiral loops cannot
be neglected and it is not hard to see that to a good approximation we
would expect the lhs of Eq.(\ref{eq:3}) to vanish when $m_\pi$ is
approximately equal to the physical $\Delta$-N mass difference. This
is indeed consistent with the lattice results.

\subsection{Other Excited States}
In terms of the application of lattice QCD to hadron spectroscopy, the
result of the previous subsection, namely that the $\Delta$ resonance
actually becomes a stable state when the pion mass exceeds the
physical $\Delta$-N mass difference, is an important result. It is far
easier to use lattice techniques to find a stable particle, which is a
true eigenstate of the QCD Hamiltonian, than to extract information
about a resonance which is not the lowest energy state with a
particular set of quantum numbers. Of course, in the case of the
$\Delta$ the situation is even better because the $\pi$-N channel must
have angular momentum one ($L = 1$) for which the minimum, non-zero
momentum on a lattice of size $Na$ is $2 \pi /Na$. Hence the $\Delta$
is stable on the lattice unless $m_\Delta < \sqrt{m_N^2+(2 \pi/Na)^2}
+ \sqrt{m_\pi^2 + (2 \pi/Na)^2}$, which is typically several hundred
MeV higher than the naive threshold, $m_N + m_\pi$.

Another case of considerable interest involving $L = 1$ is the $\rho$
meson which has a p-wave decay into two pions. Of course, for the
physical $\rho$ this decay involves momenta that are too high for the
comfortable application of effective field theory. Nevertheless, it
must be accounted for if one hopes to obtain a meaningful value of the
physical mass. It helps that the width of the $\rho$ is well known and
this provides a powerful constraint on the phenomenological treatment
of this channel~\cite{Allton:2005fb,Leinweber:2001ac}. 
In any case, for the present
discussion what matters is that the $\rho$ is actually stable on the
lattice even when $m_\rho$ is below $2 m_\pi$ and a data point at such
a low mass provides a powerful constraint on the chiral extrapolation
needed to obtain the physical mass of the $\rho$.
\begin{figure}[tp]
\includegraphics[height=12.0cm,angle=90]{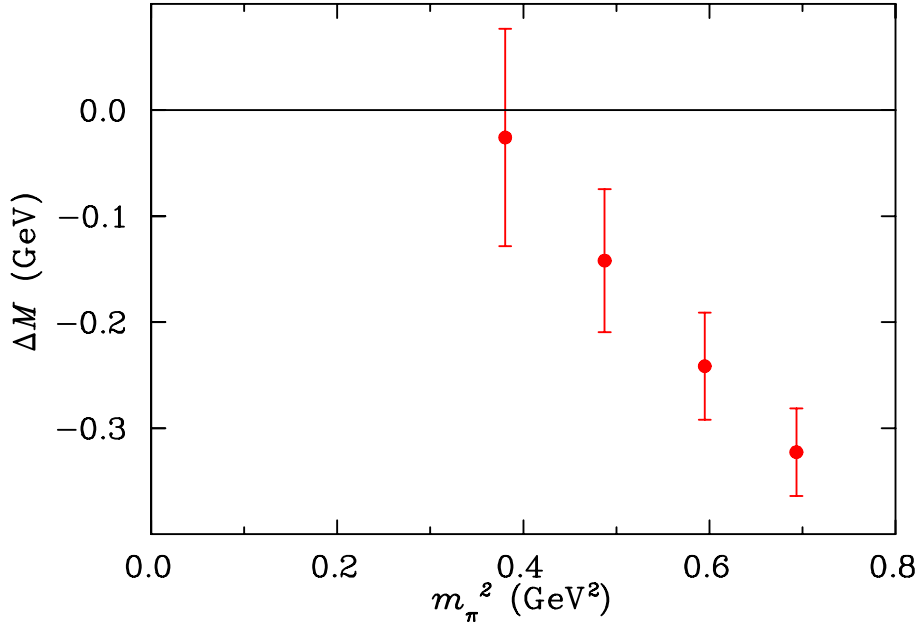}
\caption{\label{fig:1}
        Mass difference between the lowest-lying negative parity
        excited nucleon bound state,
        the $I(J^P)={1 \over 2}({1 \over 2}^-)\ N^*(1535)$,
        and the S-wave $N+\pi$ two-particle scattering state~\protect\cite{Lasscock:2005tt}.}
\end{figure}

However, the challenge of greatest interest to us here is baryon
spectroscopy and it is interesting to see whether it is more generally
true that nucleon excited states become stable as the quark mass
increases. Figure~\ref{fig:1}, from Lasscock {\it et
al.}~\cite{Lasscock:2005tt}, shows the difference between the mass of
the $1/2^{-}$ $N^*(1535)$ and its s-wave decay threshold, $m_N+m_\pi$,
as a function of pion mass. It is clearly stable for $m_\pi^2 >
0.3$--$0.4\,{\rm GeV}^2$. As in the case of the $\Delta$ this is very
close to the point where the pion mass is equal to the difference
between the physical resonance and nucleon masses ($m_\pi \sim
0.6\,{\rm GeV}$).

A similar result seems to hold for all nucleon excited
states~\cite{Sasaki:2001nf,Gockeler:2001db,Melnitchouk:2002eg,Leinweber:2004it}
which look as though they might match the corresponding experimental
state after a reasonable chiral extrapolation. As we have seen
earlier, this is what we expect as a first approximation in the case
of any nucleon excited state, given that lattice QCD shows that the
CQM seems to describe the behavior of non-perturbative QCD in the
region where the light quark masses are large ($m_\pi > 0.5$ GeV).
The key to this behavior is the rapid variation of the pion mass and
hence the corresponding threshold energy, while the energy difference
between the mass of the nucleon and the excited state varies relatively
slowly.

\section{Possible Pentaquarks}

It is difficult to imagine a more important discovery in modern strong
interaction physics should the pentaquark story have a positive
conclusion.  At the present time the experimental situation is
certainly confusing, with almost as many negative findings as positive
and some earlier announcements now contradicted with much higher
statistics~\cite{Dzierba:2004db,Hicks:2005gp}. Theoretically the
situation is just as inconclusive. It is difficult to take predictions
based on various quark models seriously for a state where we have so
little experience.  As a result there has been a particular interest
in studies based on lattice QCD, with a considerable amount of effort
already applied to the problem.

Apart from the challenge of finding an appropriate source which has a 
large amplitude for producing a pentaquark, one has the bigger challenge 
of deciding exactly what signal one is expecting to find. We saw in the 
previous section that it has been possible to study nucleon excited states 
on the lattice because they become stable as the light quark mass increases. 
Without this property it would have been a much harder job. The question 
then is what one might expect with respect to the stability of the 
pentaquark.

To investigate this question we begin with an expression for the mass of 
a pentaquark as a function of light quark mass motivated by a CQM which 
should be valid at sufficiently large $m_q$:
\be
m_5 = M_s + 4 M + H_5^{\rm hyp} \, .
\label{eq:pq1}
\ee
Here $M_s$ is the constituent mass of the strange quark, which is fixed (at 
0.565 GeV in the AccessQM) in this study of light quark mass variation. 
Using Eq.(\ref{eq:1}) we find
\be
m_5 = 2.25 +1.20 m_\pi^2 + H_5^{\rm hyp} \, .
\label{eq:pq2}
\ee
In order to compare with the corresponding threshold, in this case $K
N$, we need an expression for the kaon mass for fixed strange quark
mass. Using the Gell-Mann--Oakes--Renner relation one can write
$m_K^2 = (m_K^{(0)})^2 + m_\pi^2/2$, where $m_K^{(0)}$ is the kaon
mass in the SU(2) chiral limit.  To keep the discussion simple, we
expand the kaon mass to leading order in $m_\pi^2$
\be
m_K = m_K^{(0)} + m_\pi^2 / (4 m_K^{(0)}) = 0.485 + 0.515 m_\pi^2 + \delta_K\, .
\label{eq:K}
\ee
All higher order terms are collected into the term $\delta_K$, where to indicate the scale
at $m_\pi^2=0.4\,{\rm GeV}^2$, $\delta_K=30\,{\rm MeV}$ and at $m_\pi^2=0.8\,{\rm GeV}^2$,
$\delta_K=100\,{\rm MeV}$.

Finally, we can now compute the difference between the pentaquark
mass and the $KN$ threshold (ignoring issues of finite lattice volume
for the moment) at sufficiently large $m_q$ that we can ignore chiral loops:
\be
m_5 - m_N - m_K = 0.53 - 0.24 m_\pi^2 + H_5^{\rm hyp} - \delta_K \, .
\label{eq:diff1}
\ee

The clear difference between the situation for nucleon excited states and 
pentaquark states is that in the region of large quark mass, where the 
chiral loops for the nucleon and (presumably) the hyperfine interaction
(including chiral loops) for the pentaquark are smooth and slowly varying, 
there is not expected to be any rapid variation with $m_\pi$. In 
particular, there is no term linear in $m_\pi$ and the coefficient 
of $m_\pi^2$ is relatively small.

A plot of $m_5 - m_N - m_K$ would be expected to show only a very
smooth variation with $m_\pi$ in the region above $m_\pi^2 \sim 0.2
{\rm GeV}^2$, with typically rapid chiral variation only below this
region {\it if} this region is accessible.

The only way that the pentaquark could be stable (i.e., the lhs of
Eq.(\ref{eq:diff1}) negative) in the large mass region explored by
lattice QCD is if the hyperfine interaction in the pentaquark were
very large and attractive in that region -- an order of magnitude 
larger than the estimate reported recently in Ref.~\cite{Paris:2005sv}. 
If this were the case, then
it suggests that the only way that this state could become unbound at
the physical point  would be through the increase at low $m_\pi$ in the known, 
large chiral corrections to the
nucleon mass. These would need to cancel a good part of
the large pentaquark hyperfine
interaction at the physical point. 
In this case the interaction responsible for producing the pentaquark
is unlikely to be chiral in origin. Rather it is more likely to have
as its origin something like gluon exchange, which is expected to
exhibit a relatively smooth variation as in the CQM (typically like
$1/M$).

The other alternative, should the pentaquark exist, is that the
hyperfine interaction is relatively weak. In this case,
Eq.~(\ref{eq:diff1}) indicates that the pentaquark is likely to remain
unstable at larger quark masses. This would force one into the
situation of needing to provide a careful analysis to identify one-
and two-particle states in a finite-volume lattice simulation. To make
a connection with the physical limit, the chiral corrections must
necessarily be important, and one must consider
\be
m_5 - m_N - m_K = 0.53 - 0.24 m_\pi^2
+ \Sigma_5( {\rm chiral \, loops}) - \Sigma_N( {\rm chiral \, loops}) \, , 
\label{eq:diff}
\ee
where $H_5^{\rm hyp}$ and $\delta_K$ are assumed negligible. At the
physical point, $\Sigma_N\sim -300\,{\rm MeV}$ and thus for the pentaquark to
exist at $\sim 100\mev$ above threshold, the chiral corrections must
be of order $\sim -700\mev$. The dynamical origin of the pentaquark
in this case would therefore be chiral in nature.
\begin{figure}[tp]
\includegraphics[height=12.0cm,angle=90]{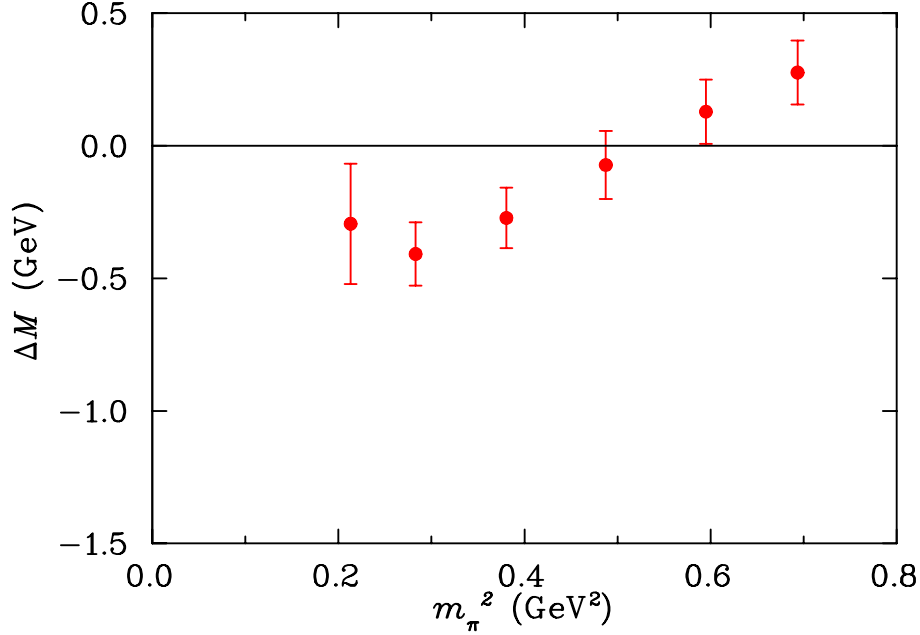}
\caption{\label{fig:2}
        Mass difference between the lowest-lying
        5-quark $I(J^P)={0}({3 \over 2}^+)$ and the $N+K$ 2-particle
        state~\protect\cite{Lasscock:2005kx}.}
\end{figure}

\section{Conclusion}
We have explored the degree to which the constituent quark model can 
quantitatively describe the behavior of baryon properties in the region of 
relatively large quark mass currently accessible to lattice QCD. In fact 
it works very well, providing further support to calls to develop modern 
constituent quark models in this mass regime and to make the connection 
with experimental data through chiral extrapolation. Of particular interest 
with respect to baryon spectroscopy is a very natural explanation of 
the reason why nucleon excited states are stable for $m_\pi$ greater than 
roughly the difference of the physical mass of the excited state and the mass 
of the nucleon.

When the same ideas are applied to the possible existence of a
pentaquark, the conclusions are somewhat different. For example, a
plot of $m_5 - m_N - m_K$ would be expected to show only a very smooth
variation with $m_\pi$ in the region above $m_\pi^2 \sim 0.2 {\rm
GeV}^2$, with typically rapid chiral variation only below this region
-- {\it if} it is accessible.  It is interesting that this is
precisely the type of behavior seen in the recent investigation of a
possible spin-3/2, positive parity pentaquark state by the CSSM-JLab
collaboration -- see Fig.~2. Clearly this needs further investigation.

Furthermore, these ideas lead us to the conclusion that, if the
signal observed in Fig.~2 holds under further study, 
the interaction responsible for producing
pentaquarks is unlikely to be chiral in origin. Rather it is more
likely to have as its origin something like gluon exchange, which is
expected to exhibit a relatively smooth variation as in the CQM
(typically like $1/M$). Finally, we note that while the slope of $\Delta M$ 
at large $m_\pi$ in Fig.~2 differs from that in Eq.~(\ref{eq:diff1}), 
this may be consistent with a $1/M$ dependence of a large, attractive 
hyperfine interaction.

\newpage 

\section*{Acknowledgments}
It is a pleasure to acknowledge many helpful conversations with
D.~Leinweber concerning the AccessQM, M.~Paris regarding quark-model
studies and B.~Lasscock and collaborators regarding the lattice simulations 
of pentaquark states.  This work was supported by
DOE contract DE-AC05-84ER40150, under which SURA operates Jefferson
Laboratory.

\end{document}